\documentclass[11pt,twoside]{article}
\usepackage{graphicx,epsfig,natbib,epstopdf}
\usepackage{CS18}
\begin{document}
%
%
\title{Stellar Spectroscopy during Exoplanet Transits: \\ Dissecting fine structure across stellar surfaces}
\author{Dainis Dravins$^{1}$, Hans-G\"{u}nter Ludwig$^{2}$, Erik Dahl\'{e}n$^{1}$, Hiva Pazira$^{1,3}$} 
\affil{$^1$Lund Observatory, Box 43, SE--22100 Lund, Sweden}
\affil{$^2$Zentrum f\"{u}r Astronomie der Universit\"{a}t Heidelberg, Landessternwarte K\"{o}nigstuhl, DE--69117 Heidelberg, Germany}
\affil{$^3$Department of Astronomy, AlbaNova University Center, SE--10691 Stockholm, Sweden}
\begin{abstract}
Differential spectroscopy during exoplanet transits permits to reconstruct spectra of small stellar surface portions that successively become hidden behind the planet.  The center-to-limb behavior of stellar line shapes, asymmetries and wavelength shifts will enable detailed tests of 3-dimensional hydrodynamic models of stellar atmospheres, such that are required for any precise determination of abundances or seismic properties.  Such models can now be computed for widely different stars but have been feasible to test in detail only for the Sun with its resolved surface structure.  Although very high quality spectra are required, already current data permit reconstructions of line profiles in the brightest transit host stars such as HD~209458 (G0 V).
\end{abstract}
\section{Context}

Three-dimensional and time-dependent hydrodynamic simulations are now established as realistic descriptions for the convective photospheres of various classes of stars, and must be applied in any more precise determinations of chemical abundances, oscillation properties, or atmospheric structure.  Such models can now be produced for stars with widely different properties, ranging from white dwarfs to red supergiants, and with all sorts of metallicities \citep{freytagetal12, magicetal13, tremblayetal13}; Fig.~1.  Although, in principle, such simulations do not have freely tunable physical parameters, their complexity implies that they must apply various physical, mathematical, and numerical approximations to be manageable.  Such models do well reproduce the details of solar spectral line profiles as well as the fine structure (granulation) observed across the solar surface, its time evolution, as well as its interaction with magnetic fields.  However, the verification (or falsification) of such models is more challenging for other stars, where surface structures cannot be spatially resolved.

\begin{figure}[ht!]
\centering
\includegraphics[width=100mm]{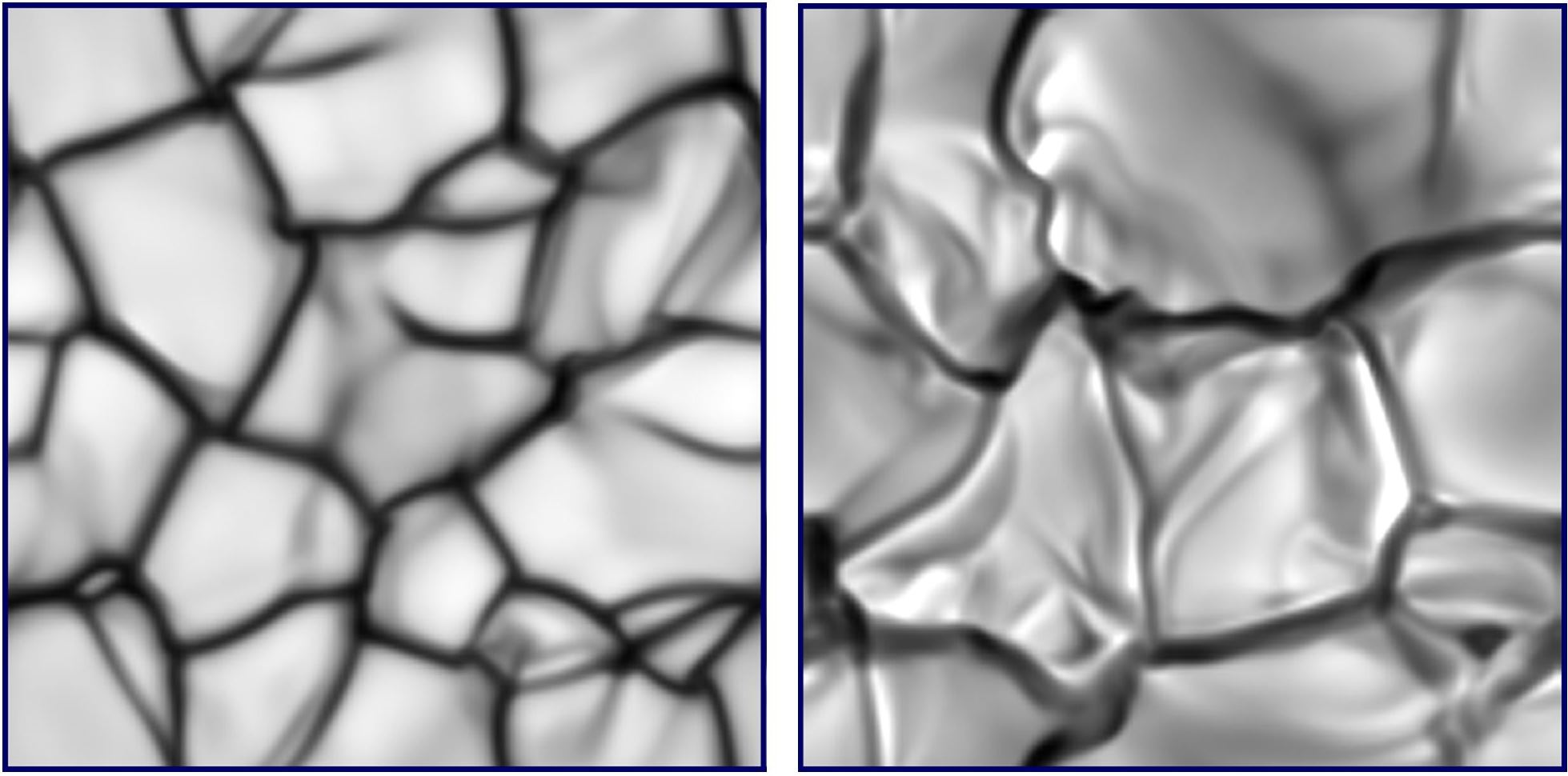}
\caption{Simulations of 3-dimensional hydrodynamics in stellar atmospheres: Examples of emergent intensity during granular evolution on the surface of a 12,000 K white dwarf (left) and a 3,800 K red giant.  The areas differ by many orders of magnitude: 7$\times$7~km$^{2}$ for the white dwarf, and 23$\times$23~R$_{\odot}$$^{2}$ for the giant. It has become possible to model widely different stars, but the observational means for verifying such simulations remain limited, except for the Sun.}
\label{overflow}
\end{figure}

Using the output from such simulations as temporally and spatially varying model atmospheres, synthetic spectral line profiles can be computed as temporal and spatial averages over the simulation sequence \citep{becketal13, holzreutersolanki13, pereiraetal13}.  Details of the atmospheric structure and dynamics are reflected in the exact profiles of photospheric absorption lines and in their center-to-limb variations.  Line asymmetries and wavelength shifts originate from the statistical bias of an often greater number of [spectrally blueshifted] photons from hot and rising surface elements (granules) than from the cooler and sinking ones. Center-to-limb changes depend on the relative amplitudes of horizontal and vertical velocities, whose Doppler shifts contribute differently to the line broadening.  These changes also depend on the amount of `corrugation' across the stellar surfaces: if those are `smooth' (in the optical-depth sense), convective blueshifts should decrease from disk center towards the limb, since vertical convective velocities become perpendicular to the line of sight, and the horizontal velocities that contribute Doppler shifts appear symmetric.  However, stars with `hills' and `valleys' should show an increasing blueshift towards the limb, where one will predominantly see the approaching (thus blueshifted) velocities on the slopes of the `hills' facing the observer.   The depth dependence of atmospheric properties causes dependences on parameters such as the line's oscillator strength, its excitation potential, ionization level, and the wavelength region.  Obviously, such types of spectral data could open up stellar surface structure to rather detailed observational study \citep{dravinsnordlund90, dravinsetal05}.

\section{Stellar physics from exoplanet transits}

Exoplanet transits offer a possibility to disentangle spectra from small portions across the stellar surface.  During transit, an exoplanet hides successive segments of the stellar disk and differential spectroscopy between epochs outside transit, and those during different transit phases, provide spectra of those particular surface segments that were then hidden behind the planet (Fig.~2).  If the planet would happen to cross a starspot, even spatially resolved spectra (with their magnetic signatures) of such stellar surface features could become attainable, provided sufficient spectral fidelity can be reached.  Knowing the precise `background' stellar spectrum at different disk positions, also measurements of exoplanet atmospheres during transit would be better constrained.

\begin{figure}[ht!]
\centering
\includegraphics[width=110mm]{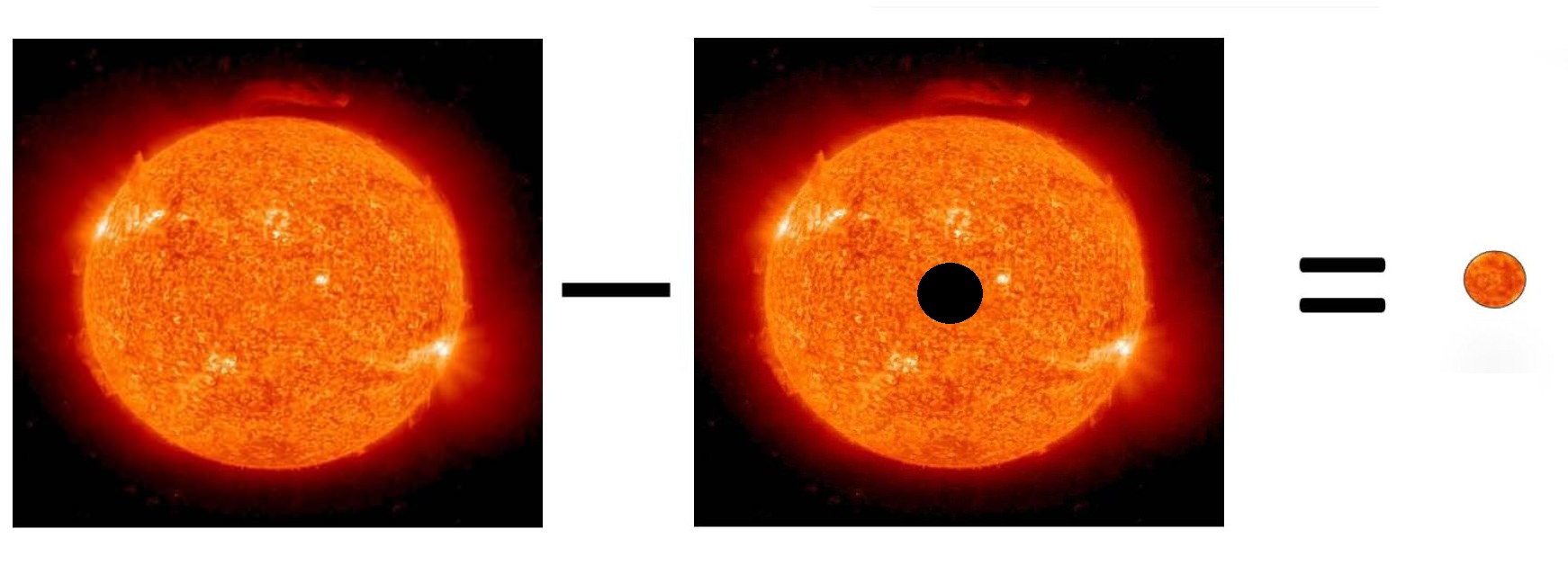}
\caption{The difference between the flux recorded outside, and during exoplanet transit, yields the signal of the temporarily ‘hidden’ stellar surface segment.}
\label{overflow}
\end{figure}

Although the method is straightforward in principle, it is observationally very challenging since exoplanets cover only a tiny fraction of the stellar disk (no more than  $\sim$1\% for main-sequence stars).  If a desired signal-to-noise in the reconstructed spectrum would be on order S/N = 100, say, extracted from only $\sim$1\% of the total stellar signal, that would require a S/N in the latter on the order of 10,000 or more.  This may appear daunting, but the method actually turns out to be (at least marginally) feasible for the brightest exoplanet host stars already with current spectrometers (e.g., UVES on VLT).  In the near future one can expect it to become more potent; both because ongoing exoplanet surveys are likely to find brighter planet hosts, and because forthcoming higher-performance spectrometers (e.g., PEPSI on LBT or ESPRESSO on VLT) will enable such studies for also fainter and rarer stellar types.

\section{Synthetic spectral lines from 3-D modeling}

Examples of synthetic spectral lines computed from CO{$^5$}BOLD models \citep{freytagetal12} illustrate what types of signatures to expect (Figs.~3-4).  The simulations are made for a certain spatial area on the star, and run for sufficiently many timesteps, so that the output is statistically stable.  Spectral line profiles are then computed for each spatial location across the simulation area, for each timestep, and for different inclination angles to produce representative spatial and temporal averages for different center-to-limb positions.  Since such spectroscopic measures are insensitive to (modest amounts of) spatial smearing, they can be compared to observations with modest spatial resolution; already spectra from a small number of center-to-limb positions will provide unique information.

\begin{figure}[ht!]
\centering
\includegraphics[width=67mm]{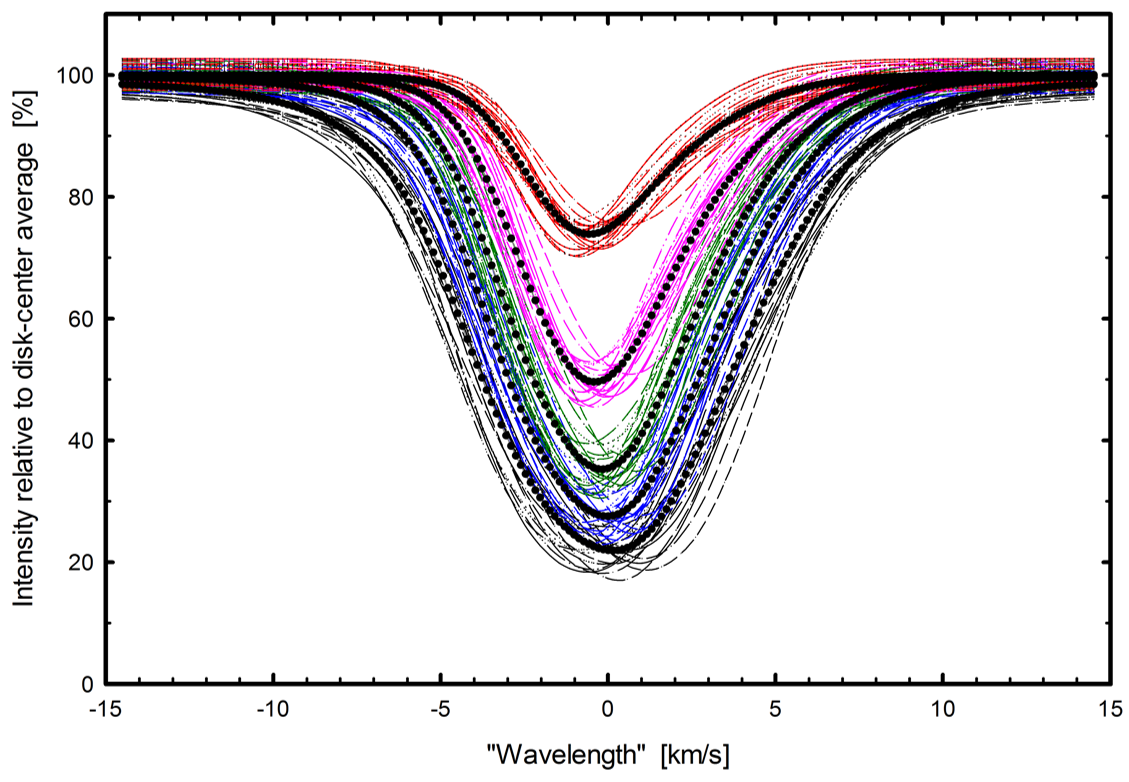}
\caption{Synthetic Fe I lines at stellar disk center ($\mu$ = cos$~\theta$ = 1.0) from a CO\,$^{\it{5}}\!$BOLD model of a giant star with approximate spectral type K0 III (T$_{e\!f\!f}$= 5000 K, log g [cgs] = 2.5).  Thin lines are averages over the spatial simulation area for each of 20 timesteps; bold curves are their temporal averages.  Five different line strengths are shown; $\lambda$ = 620 nm, $\chi$ = 3 eV. }
\label{overflow}
\end{figure}

Fig.~3 illustrates some characteristic features of spectral lines formed in convective atmospheres. The lines generally become asymmetric and wavelength-shifted and these effects depend on line-strength; e.g., the weaker lines display a stronger convective blueshift.  Fig.~4 shows other dependences, now also including that on excitation potential, and the position across the stellar disk.  Clearly, such data encode a wealth of information.

\begin{figure}[ht!]
\centering
\includegraphics[width=132mm]{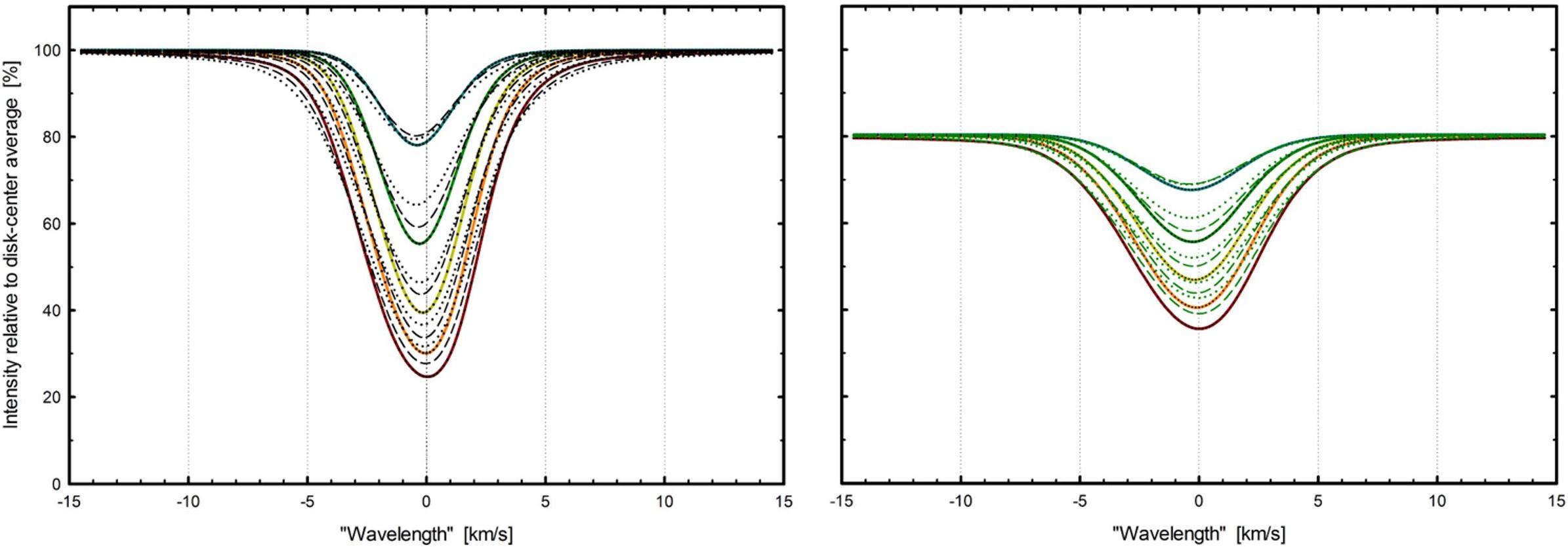}
\caption{Synthetic time-averaged photospheric Fe I lines for a solar model.  Left: Disk center ($\mu$ = cos$~\theta$ = 1.0); five different line strengths, each for the three excitation potentials $\chi$ = 1, 3, and 5 eV.  Right: Same for disk position $\mu$ = cos$~\theta$ = 0.59.  The decreased intensity reflects the limb darkening at $\lambda$=620 nm while increased linewidths reflect that average horizontal velocities are greater than vertical ones, contributing more Doppler broadening towards the limb.}
\label{overflow}
\end{figure}

\section{Synthetic spectral lines across stellar disks}

We now examine what types of spectral-line signatures one may expect across stellar disks.  Fig.~5 shows a center-to-limb sequence for synthetic line profiles in a main-sequence star with solar metallicity, and T$_{e\!f\!f}$ = 6800 K, including the line bisectors (on an expanded scale).  Close to disk center, there is a sudden change in the bisector shape near continuum intensity, the curve suddenly turning to the blue, i.e. a `blueward hook'.  This has been previously observed for the F-type star Procyon and theoretically traced to the extended Lorentzian wings of the stronger, saturated, and blueshifted line components.  Their contribution in one flank of the spatially averaged line also affects the intensity in the opposite flank, in contrast to Gaussian-like components, whose absorption disappears over a short wavelength distance.  The steeper temperature gradients in the rising and blueshifted granules produce stronger absorption lines, which therefore tend to first saturate and develop Lorentzian damping wings in those spatial locations; \citet{allendeetal02}.  Such signatures thus reveal differences in line-formation conditions in different inhomogeneities across the stellar surface.

\begin{figure}[ht!]
\centering
\includegraphics[width=132mm]{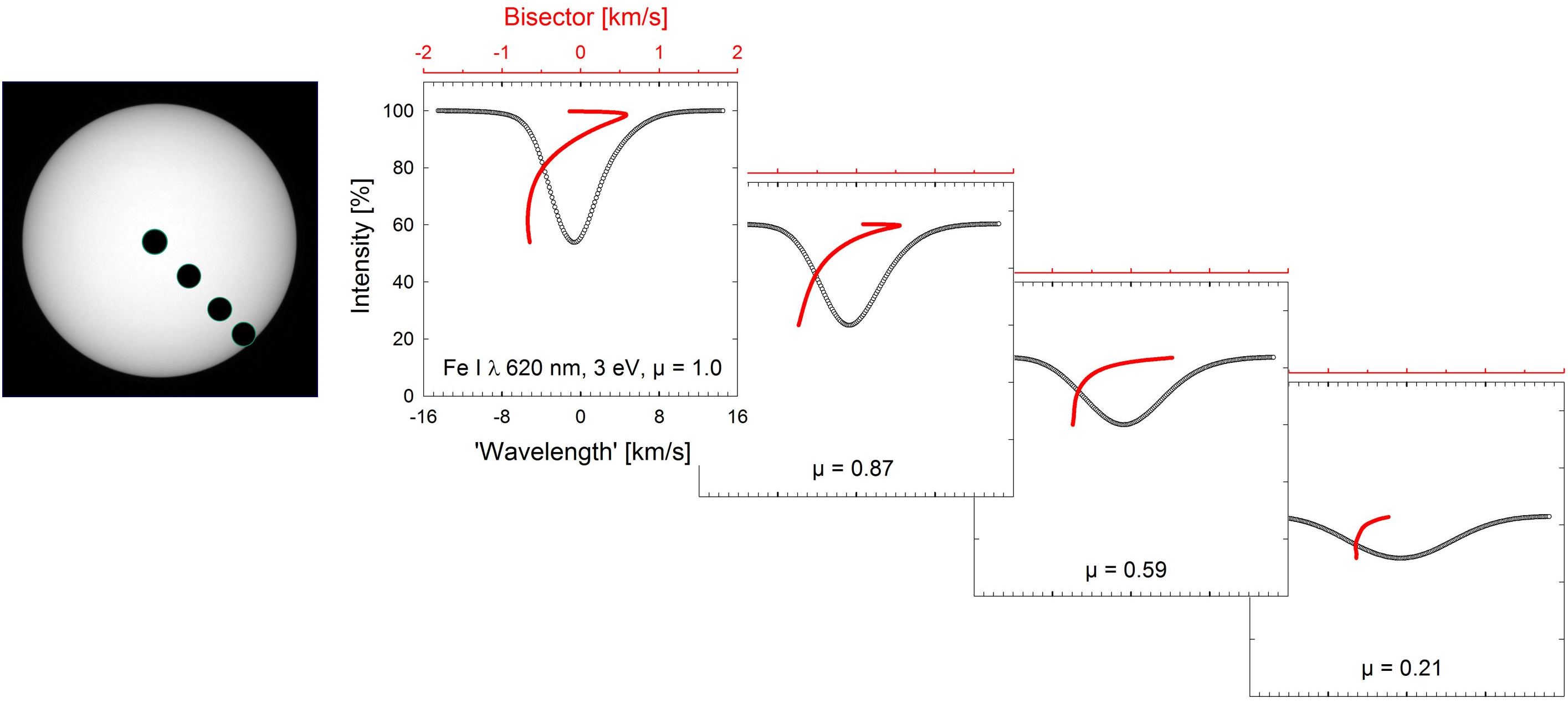}
\caption{Sequence of synthetic Fe I line profiles across the stellar disk, computed from a CO\,$^{\it{5}}\!$BOLD model of a main-sequence star of solar metallicity, T$_{e\!f\!f}$ = 6800 K.  The line asymmetries and wavelength shifts are shown by the bisectors.}
\label{overflow}
\end{figure}

\section{Simulated line changes during exoplanet transit}

As already noted, the spectrophotometric fidelity required for the corresponding observations is quite challenging.  An examination of various options in retrieving spatially resolved stellar spectra suggests that a ratio method appears to be the one least sensitive to systematic and random noise.     
Fig.~6 shows the expected line-profile ratios during (half of) an exoplanet transit across the stellar equator.  The red solid curves show the ratios of line profiles relative to that outside transit.  This simulation sequence from a CO\,$^5$BOLD model is for an Fe I line ($\lambda$ 620 nm, $\chi$ = 3 eV) during the transit by a `bloated' Jupiter-size exoplanet moving in a prograde orbit across the stellar equator, covering 2\% of this main-sequence star with solar metallicity, T$_{e\!f\!f}$  = 6300 K, rotating with $V$ = 5 km~s$^{-1}$.

\begin{figure}[ht!]
\centering
\includegraphics[width=100mm]{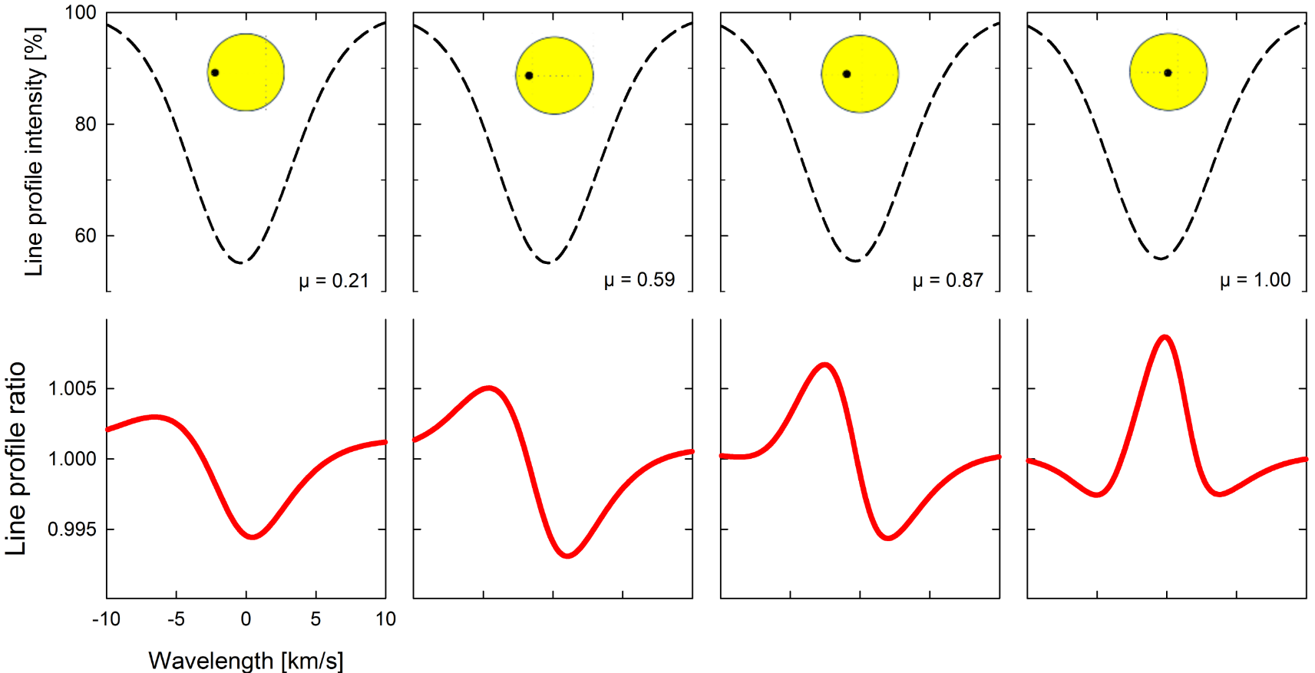}
\caption{Simulated line changes during the first half of an exoplanet transit across the equator of a  T$_{e\!f\!f}$ = 6800 K main-sequence star.  Red: Ratios of line profiles relative to the profile outside transit.  }
\label{overflow}
\end{figure}

\subsection{Simulated Rossiter-McLaughlin Effect}

There are other measures that can be observed during a transit, e.g., the change of the star's apparent radial velocity.  This `Rossiter-McLaughlin effect' occurs because the exoplanet successively hides different portions of a rotating star with locally either blue- or redshifted Doppler components, tipping the disk-averaged wavelength slightly towards longer or shorter values. 

Fig.~7 shows the Rossiter-McLaughlin effect simulated for the same model as that for Fig.~6.  The concept of `wavelength' (not unique for asymmetric spectral lines) is here defined as fits of Gaussian functions to the synthetic line profiles.  The values are negative (i.e., shorter than laboratory values) because of the convective blueshift.  It can also be noted that the curves before and after mid-transit ($\mu$ = cos$~\theta$ = 0.21, 0.59, 0.87) are not exact mirror images of one another due to intrinsic stellar line asymmetries. 

\begin{figure}[ht!]
\centering
\includegraphics[width=60mm]{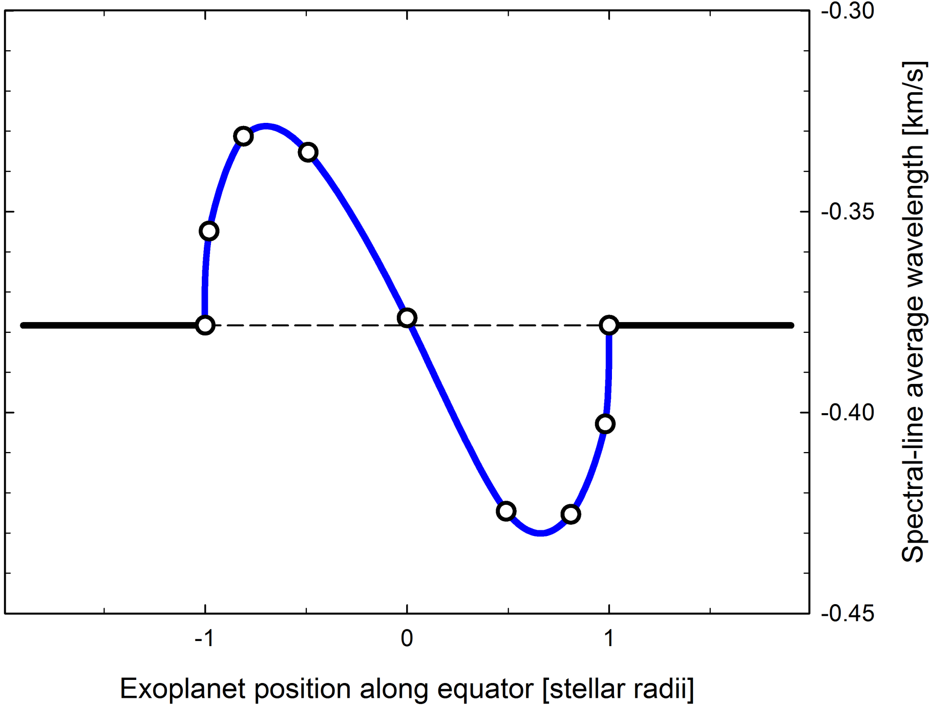}
\caption{Simulated Rossiter-McLaughlin effect for a T$_{e\!f\!f}$ = 6800 K star rotating with equatorial velocity = 5 km~s$^{-1}$, and a transiting planet covering 2\% of she stellar disk.  The spectral-line `wavelengths' were here obtained as Gaussian fits to the synthetic (and asymmetric) line profiles.}
\label{overflow}
\end{figure}

\subsubsection{Exoplanet transit geometry}

The geometry of the exoplanet transit must be known in order to fully interpret any observed line-profile variations (only in rare cases will the transit occur along the stellar equator).  This can be constrained through the Rossiter-McLaughlin effect, enabling the planet's position and path across the stellar disk to be determined; e.g., \citet{Torresetal08}.

\section{Observations with current facilities}

The demanding signal-to-noise requirements (e.g., the line-ratio amplitudes seen in Fig.~6 have amplitudes of only 0.5\%) limit usable data to the very highest-fidelity spectra from high-resolution spectrometers at very large telescopes.  Although very many transiting planets are known, realistic targets include only such where the exoplanet is large (Jupiter-size or bigger) and the star among the brightest hosts found so far.  Of course, these are the very objects of which extensive observations have been carried out in order to characterize the exoplanet and its atmosphere, and where numerous spectra are available in observatory archives.  For some dozen such candidates, many hundreds of archive spectra were retrieved from several different observatories and examined for their suitability within the present project.  The particular data shown in the following figures originate from one observation night with the UVES spectrometer \citep{dekker00} on the VLT Kueyen telescope of ESO on Paranal as part of an investigation of the atmosphere of the exoplanet HD~209458b, for which other results were documented in \citet{albrechtetal09} and \citet{snellenetal11}. The nominal random signal-to-noise values, as computed by the ESO data pipeline, reached values exceeding 500 in the centers of the best-exposed echelle orders. Exposure times were 400 seconds during and outside the three-hour transit, with nominal spectral resolution $\lambda$/$\Delta\lambda$ $\approx$~65,000. This is a large planet (`bloated hot Jupiter', with R = 1.38 R$_{J\!u\!p}$), covering some 1.5~\% of the stellar surface, close to the maximum for a solar-type main-sequence star.  Exoplanets are not expected to get much larger, but some might well be found around smaller red (or even white) dwarfs, which would make these studies easier (but at the cost of a lower relative spatial resolution on the star).    

Fig.~8 shows preliminary data for reconstructed H$\beta$ line profiles on various positions across the disk of HD~209458.  The star is close to the Sun in spectral type (G0~V, sometimes also classified as F9~V; T$_{e\!f\!f}$ = 6100 K, log~g [cgs] = 4.50; [Fe/H] = 0; $V_{rot}$ = 4.5 km~s$^{-1}$, and sin~$i$ = 1 if the star rotates in the same plane as its transiting planet).  The solid curve shows the spectrum outside transit, obtained as an average over numerous recordings, and the dashed curves show the extracted spectrum behind the exoplanet during various phases of the transit: longer-dashed during earlier parts, and shorter-dashed curves from later parts of the transit.  As can be seen, the line gradually shifts towards longer wavelengths, as appropriate for a planet in a prograde orbit.  In a sense, this is the spatially resolved version of the Rossiter-McLaughlin effect, directly revealing the magnitude of the stellar rotational velocity vector.  

\begin{figure}[ht!]
\centering
\includegraphics[width=90mm]{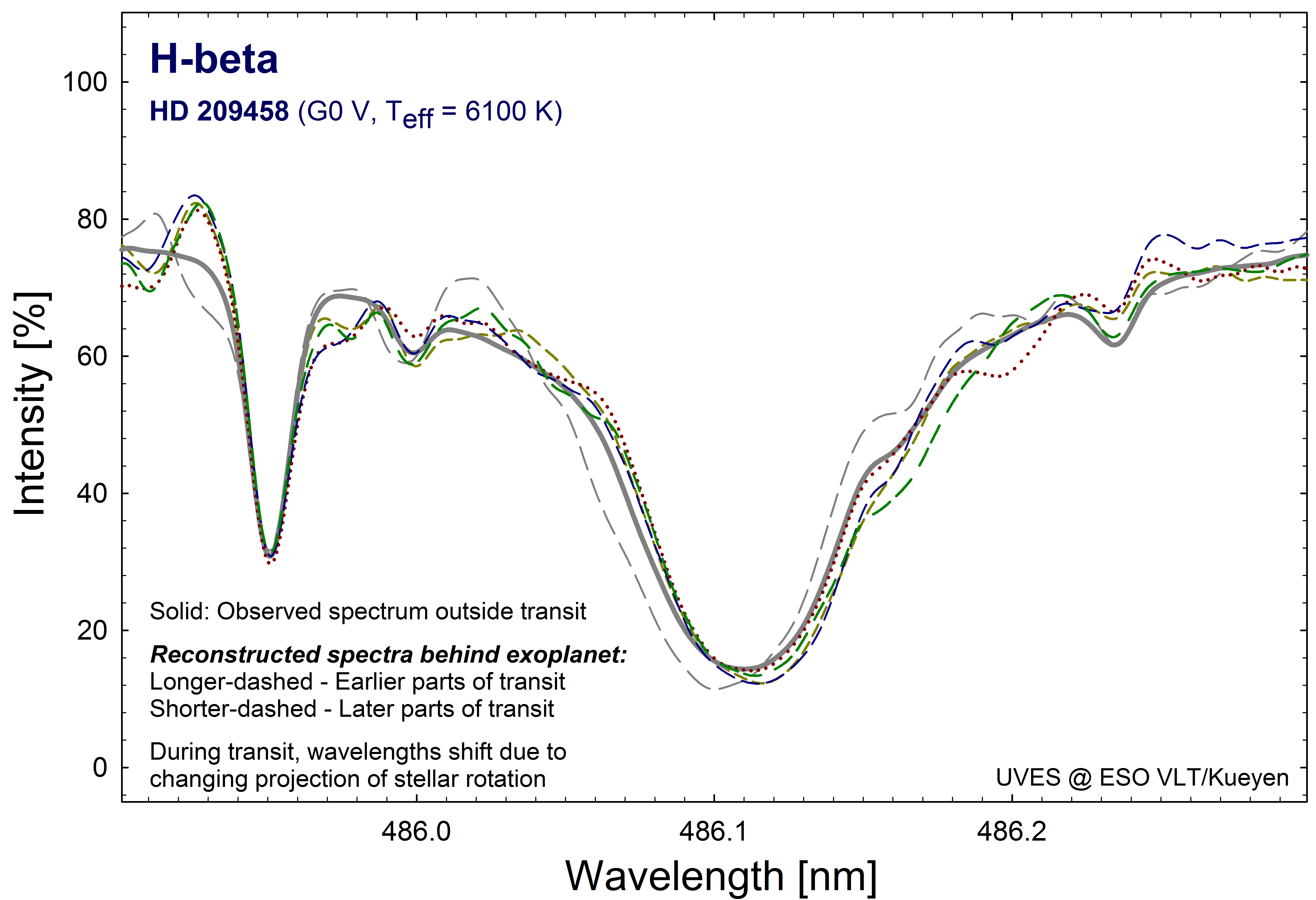}
\caption{Reconstructed H$\beta$ profiles for different positions across the disk of HD~209458 using spectra recorded with the UVES spectrometer at ESO/VLT.}
\label{overflow}
\end{figure}

\begin{figure}[ht!]
\centering
\includegraphics[width=90mm]{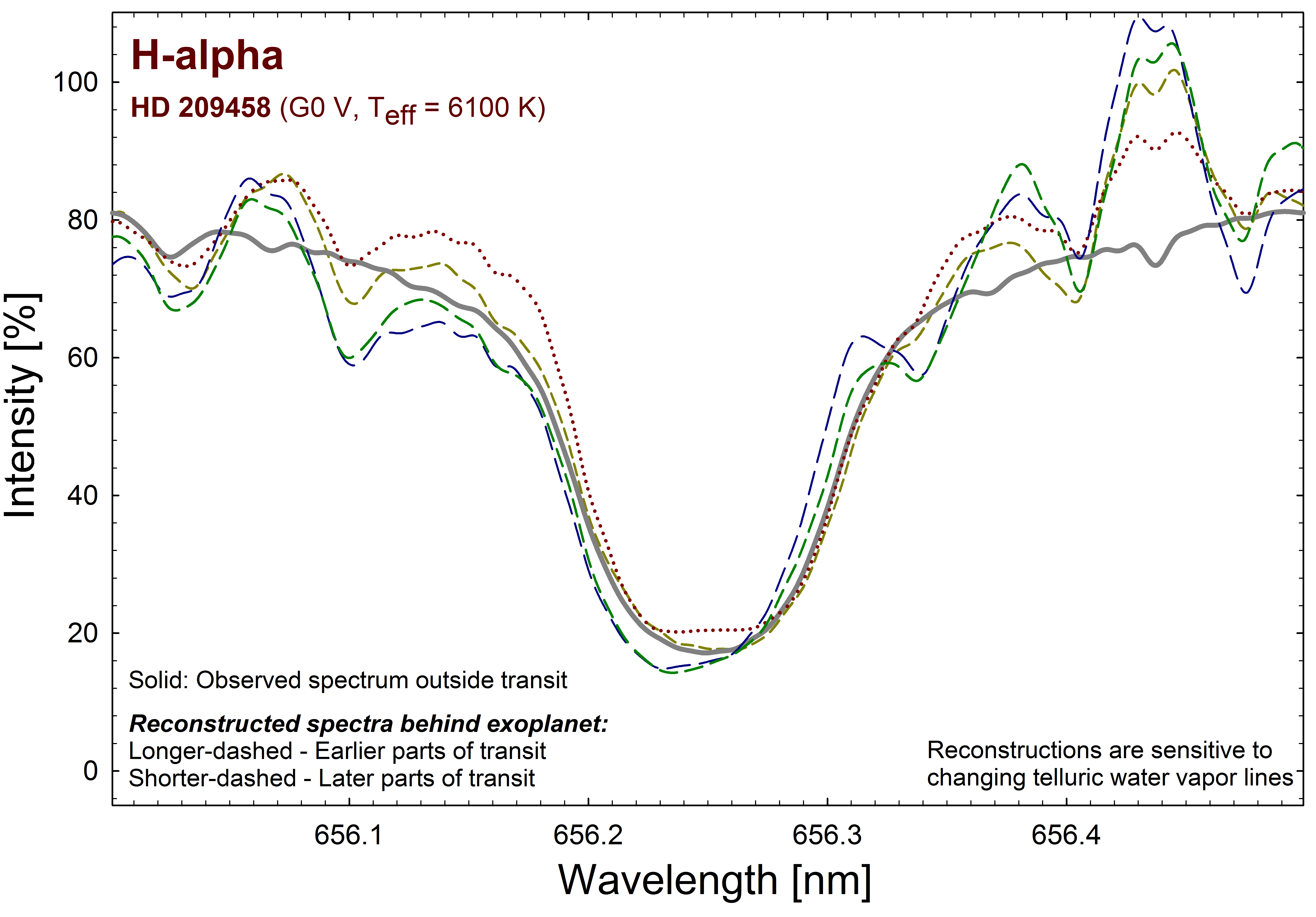}
\caption{Reconstructed H$\alpha$ profiles across the disk of HD~209458.  The apparent `emission' in the longward part is an artifact from varying  telluric water vapor lines during the hours of transit and outside it.}
\label{overflow}
\end{figure}

Fig.~9 shows analogous reconstructions around the H$\alpha$ line.  Besides the same sequence of gradually shifting signatures of stellar rotation, this illustrates another challenge in such reconstructions.  Around $\lambda$~656.5 nm, some spurious `emission' appears.  This was identifies as an artifact originating from variations in the strength of telluric water vapor lines during the several hours of transit and the period outside it. The reconstruction obtains that line profile, which is required to make its intensity-weighted summation with the observed transit profile, equal to the profile from outside transit.  Already a small change of telluric line strengths between the transit and out-of-transit epochs implies that the deduced spectrum from behind the exoplanet must carry all this change.  Since it carries only a small weight corresponding to its tiny area coverage, this spectrum must then be significantly modified to account for it all.  Even if water vapor effects are most prominent in the red part of the spectrum (like here around H$\alpha$), this exemplifies the need to control systematic errors.

Strong chromospheric lines such as these Balmer ones do not obtain their widths from photospheric motions, and their relative center-to-limb changes are expected to be rather smaller than for photospheric ones.  A further aspect is that these strong lines are also those where some contribution from the extended atmosphere of the exoplanet may be expected.  However, their large widths makes them less challenging to reconstruct, which is the reason why they were chosen for these first data samples.  A somewhat narrower line is represented by the Na~I~D doublet (also studied in the context of the exoplanet atmosphere).  Its greater intensity gradients makes it more susceptible to noise, with the photometric limitations illustrated in Figures 10 and 11.  The average profiles for earlier and later parts of the transit do again clearly show the stellar rotation signature with amplitude on order 5 km~s$^{-1}$, consistent with rotational estimates from full-disk photospheric line broadening. The sequences in Figure 11 of the ratios of successive 400-second exposures to the reference spectrum from outside transit illustrate how line profile changes can be followed during transit well above the noise level seen in out-of-transit spectra, however not by a very large margin.  The pattern of stable differences in the transit spectra are believed to originate from differences in telluric water-vapor lines between the hours of transit and the times of recording out-of-transit spectra.

\begin{figure}[ht!]
\centering
\includegraphics[width=90mm]{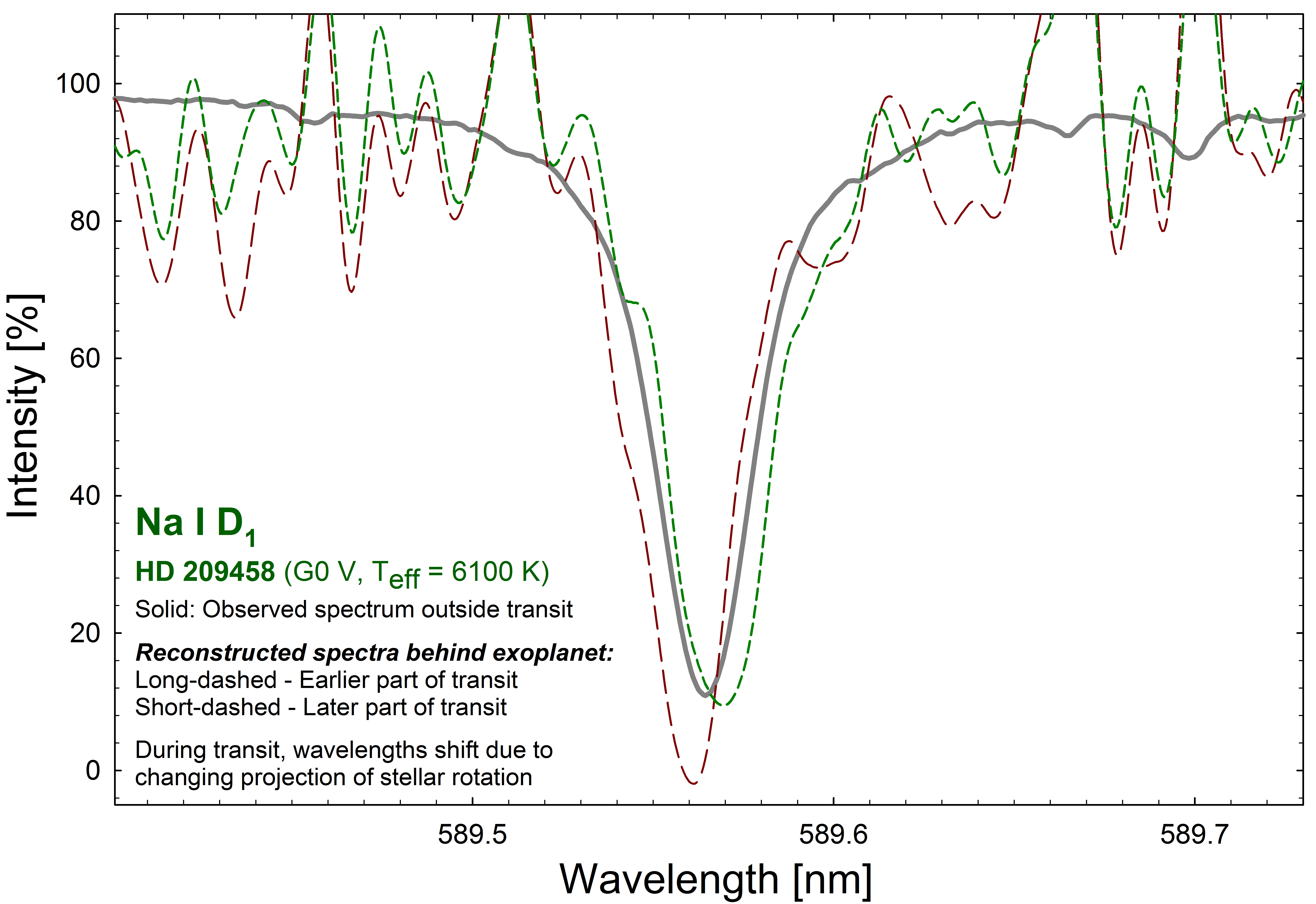}
\caption{Reconstructed Na~I~D$_1$ profiles across the disk of HD~209458.  The noise level in current UVES spectra starts to become marginal for these narrower Na~I~D lines.  Shown are average profiles for earlier and later parts of the transit.}
\label{overflow}
\end{figure}

\begin{figure}[ht!]
\centering
\includegraphics[width=90mm]{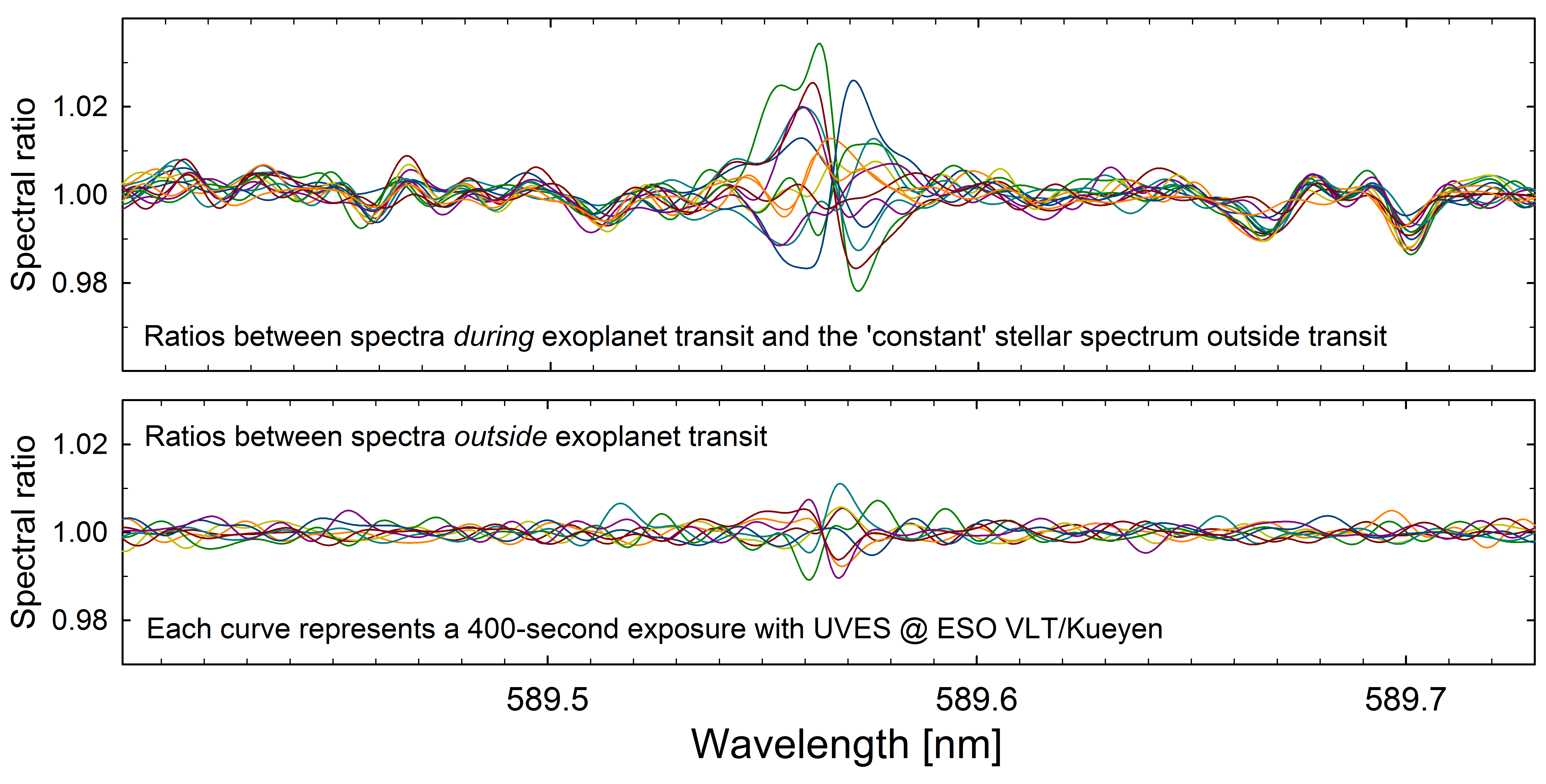}
\caption{Spectral fidelity around Na I D$_1$: Successive exposures compared to the reference spectrum from outside transit.  Line profile changes can be followed during the exoplanet transit (top) with the noise level illustrated by analogous data outside transit (bottom).  Changes during the observing night in telluric line-strengths are believed to cause the stable ratio patterns outside the Na I line.}
\label{overflow}
\end{figure}

These examples show that: (i) Already with data from existing facilities, it is feasible to reconstruct stellar line profiles from tiny portions of a stellar disk although (ii) for {\it{individual}} single lines, observed during only one transit, the very high signal-to-noise level required permits this to be done only for the broadest and strongest chromospheric ones.  At these S/N levels, also close attention has to paid to possible sources of systematics, such as variable telluric absorption in some spectral regions, the stability of spectrometer calibration, of its software routines, and other. 

The longer-term aim is to retrieve narrow photospheric line profiles, such as shown in Figure 5.  Although, with current facilities and current stellar targets, this appears not practical for {\it{individual}} spectral lines, it may well be possible taking advantage of the multitude of physically similar ones (an option not available for the strong but few and mutually different chromospheric lines).  However, already from just Fe I, the visual spectrum of solar-type stars contains on the order of 1000 measurable photospheric lines, of which perhaps one half can be classified as reasonably `unblended'.  These can then be subdivided into perhaps five groups of differently strong ones, with $\sim$100 lines per group.  For lines with similar formation conditions, their profiles are shaped in a similar fashion and the information content in their line profiles is basically redundant.  Thus, a suitable averaging of lines in such groups must be able to increase the data amount by a factor of $\sim$100, correspondingly reducing the noise level.  There is a requirement of sufficiently high spectral resolution but not necessarily of a very small stellar rotational velocity.  Even if the lines appear rotationally broadened and blended in the integrated disk spectrum, their spatially resolved reconstructions will have no such broadening.  Such averaging over numerous -- different but physically similar -- lines has been made for the solar spectrum \citep{dravins08}, permitting to identify also subtle patterns of line asymmetries, otherwise hidden in noise for individual lines.  Work in this direction is currently in progress.

\section{Future observations}

As demonstrated above, retrieval of stellar spectral lines from small portions across stellar disks is feasible already with current facilities, and for current stellar targets.  Both of these parameters are about to change in the near future, likely offering much improved possibilities for spatially resolved stellar spectroscopy.

One limiting parameter comes from the photometric noise that can be reached with existing telescopes and instruments.  The data shown above originated from observations with the ESO 8.2 m VLT Kueyen telescope, with its UVES spectrometer entrance slit opened up to 0.5 arcseconds to maximize photometric precision at the cost of spectral resolution.  Such a compromise was dictated by the limited brightness of the target star, HD~209458, of visual magnitude V = 7.6.  Although brighter transiting planet hosts are currently not known (e.g., HD~189733, K0 V, is of a similar magnitude), numerous surveys are in progress or being planned, both from ground and from space, to survey brighter stars for possible exoplanet transits.  This is a highest-priority activity for exoplanet studies since the possibilities for exoplanet characterization likewise are strongly dependent on the host star brightness.  Given the known statistics of exoplanet occurrence, it is highly likely that suitable transiting planets will soon be found around also brighter hosts.  Once a target star of visual magnitude V = 5, say, is found, that will improve the signal by one order of magnitude.

Another limiting parameter comes from the telescope + spectrometer combination.  Here, very promising developments are in progress \citep{dravins10}.  The PEPSI spectrometer \citep{strassmeieretal04, strassmeieretal08}  is in the process of becoming operational at the Large Binocular Telescope with its two 8.4 m apertures.  Offering extended spectral coverage with resolutions $\lambda$/$\Delta\lambda$ up to 320,000, this has fully sufficient performance to reveal photospheric line asymmetries and other signatures of stellar photospheric structure and dynamics of convection.  Also in progress is the ESPRESSO spectrometer for the combined focus of the four VLT unit telescopes of ESO on Paranal \citep{pepeetal13, pepeetal14}.  Observing time for such studies is virtually guaranteed since the detailed spectroscopy of exoplanets during transit of bright stars is also one of the highest-priority projects of exoplanet research, and the data required for stellar analyses will be obtained concurrently.  In the somewhat more distant future, we can look forward to the proposed HIRES instrument at E-ELT, the European Extremely Large Telescope \citep{maiolinoetal13}.  This may not reach higher spectral resolution that its predecessors, but the order-of-magnitude increase in the telescope's collecting area should enable to reach also fainter and rarer stars, perhaps metal-poor, chemically peculiar and magnetically special ones.

%
%

\acknowledgments{This study used data obtained from the ESO Science Archive Facility, originating from observations made with ESO Telescopes at the La Silla Paranal Observatory under program ID: 077.C-0379(A) by Snellen, Collier, Cameron, and Horne.  At Lund Observatory, contributions to the examination of archival spectra from different observatories were made also by Tiphaine Lagadec and Joel Wallenius.  HGL acknowledges financial support by the Sonderforschungsbereich SFB881 ``The Milky Way System'' (subprojects A4 and A5) of the German Research Foundation (DFG).}


\normalsize

\begin{references}

\bibitem[Albrecht et al.(2009)]{albrechtetal09} Albrecht, S., Snellen, I., de Mooij, E., \& Le Poole, R.\ 2009, IAU Symp., 253, 520 
\bibitem[Allende Prieto et al.(2002)]{allendeetal02} Allende Prieto, C., Asplund, M., Garc{\'\i}a L\'{o}pez, R.~J., \& Lambert, D.~L.\ 2002, \apj, 567, 544
\bibitem[Beck et al.(2013)]{becketal13} Beck, C., Fabbian, D., Moreno-Insertis, F., Puschmann, K.~G., \& Rezaei, R.\ 2013, \aap, 557, A109; corrigendum \aap, 559, 1 
\bibitem[Dekker et al.(2000)]{dekker00} Dekker, H., D'Odorico, S., Kaufer, A., Delabre, B., \& Kotzlowski, H.\ 2000, Proc.\ SPIE 4008, 534
\bibitem[Dravins(2008)]{dravins08} Dravins, D.\ 2008, \aap, 492, 199 
\bibitem[Dravins(2010)]{dravins10} Dravins, D.\ 2010, Astron.\ Nachr., 331, 535
\bibitem[Dravins \& Nordlund(1990)]{dravinsnordlund90} Dravins, D., \& Nordlund, {\AA}.\ 1990, \aap, 228, 184
\bibitem[Dravins et al.(2005)]{dravinsetal05} Dravins, D., Lindegren, L., Ludwig, H.-G., \& Madsen, S.\ 2005, in ESA SP-560, 13th Cambridge Workshop Cool Stars, ed. F.~Favata et al., 113 
\bibitem[Freytag et al.(2012)]{freytagetal12} Freytag, B., Steffen, M., Ludwig, H.-G., et al.\ 2012, J.\ Comp.\ Phys., 231, 919
\bibitem[Holzreuter \& Solanki(2013)]{holzreutersolanki13} Holzreuter, R., \& Solanki, S.~K.\ 2013, \aap, 558, A20 
\bibitem[Magic et al.(2013)]{magicetal13} Magic, Z., Collet, R., Asplund, M., et al.\ 2013, \aap, 557, A26 
\bibitem[Maiolino et al.(2013)]{maiolinoetal13} Maiolino, R., Haehnelt, M., Murphy, M.~T., et al.\ 2013, arXiv:1310.3163 
\bibitem[Pepe et al.(2013)]{pepeetal13} Pepe, F., Cristiani, S., Rebolo, R., et al.\ 2013, ESO Messenger, 153, 6  
\bibitem[Pepe et al.(2014)]{pepeetal14} Pepe, F., Molaro, P., Cristiani, S., et al.\ 2014, Astron.\ Nachr., 335, 8 
\bibitem[Pereira et al.(2013)]{pereiraetal13} Pereira, T.~M.~D., Asplund, M., Collet, R., et al.\ 2013, \aap, 554, A118 
\bibitem[Snellen et al.(2011)]{snellenetal11} Snellen, I., Albrecht, S., de Mooij, E., \& Poole, R.~L.\ 2011, Astron.\ Soc.\ Pacific Conf.\ Ser., 450, 39
\bibitem[Strassmeier et al.(2004)]{strassmeieretal04} Strassmeier, K.\ G., Pallavicini, R., Rice, J.\ B., \& Andersen, M.\ I.\ 2004, Astron.\ Nachr., 325, 278
\bibitem[Strassmeier et al.(2008)]{strassmeieretal08} Strassmeier, K.~G., Woche, M., Ilyin, I., et al.\ 2008, Proc. SPIE 7014, 70140N
\bibitem[Torres et al.(2008)]{Torresetal08} Torres, G., Winn, J.~N., \& Holman, M.~J.\ 2008, \apj, 677, 1324 
\bibitem[Tremblay et al.(2013)]{tremblayetal13} Tremblay, P.-E., Ludwig, H.-G., Freytag, B., Steffen, M., \& Caffau, E.\ 2013, \aap, 557, A7 

\end{references}
\end{document}